\documentclass[aps,pre,twocolumn,superscriptaddress,showpacs]{revtex4-1}
\usepackage[dvips]{graphicx}
\usepackage{amsmath}

\usepackage[addmarkup=uline]{changes}
\definechangesauthor[color=blue]{SY}

\begin{document}

\title{Feynman Propagator for a System of Interacting Scalar Particles in
the Fokker Theory}

\author{Natalia Gorobey}
\affiliation{Peter the Great Saint Petersburg Polytechnic University, Polytekhnicheskaya
29, 195251, St. Petersburg, Russia}

\author{Alexander Lukyanenko}\email{alex.lukyan@mail.ru}
\affiliation{Peter the Great Saint Petersburg Polytechnic University, Polytekhnicheskaya
29, 195251, St. Petersburg, Russia}

\author{A. V. Goltsev}
\affiliation{Ioffe Physical- Technical Institute, Polytekhnicheskaya
26, 195251, St. Petersburg, Russia}

\begin{abstract}
A generalized canonical formulation of the theory of the electromagnetic Fokker
interaction for a system of two particles is proposed. The functional
integral on the generalized phase space is defined as the initial one in
quantum theory. After integration over the momenta, the measure of
integration in the generalized configuration space of world particle lines
is determined. The problem of dynamic interpretation of the Feynman
propagator for a system of particles with independent proper time parameters
is discussed. A modification of the propagator is proposed, in which the
role of independent time parameters is taken by the time
coordinates of the particles in Minkowski space.
\end{abstract}


\maketitle





\section{\textbf{INTRODUCTION}}

Wheeler and Feynman \cite{1,2} formulated classical electrodynamics without
the concept of electromagnetic field in terms of the Fokker principle of
least action \cite{3}. For the simplest system of two particles, the Fokker
action has the form $\left( c=1\right) $:
\begin{eqnarray}
I_{F} &=&\frac{1}{2}\int_{0}^{S_{1}}ds_{1}\left( \overset{\cdot }{x}%
_{1}^{2}+m_{1}^{2}\right) +\frac{1}{2}\int_{0}^{S_{2}}ds_{2}\left( \overset{%
\cdot }{x}_{2}^{2}+m_{2}^{2}\right)  \notag \\
&&+e_{1}e_{2}\int_{0}^{S_{1}}ds_{1}\int_{0}^{S_{2}}ds_{2}\delta \left(
s_{12}^{2}\right) \overset{\cdot }{x}_{1}x_{2}^{\prime }.  \label{1}
\end{eqnarray}%
Here, the world lines of particles are parametrized by the intrinsic time of
each particle, the dot and dash denote the derivatives with respect to the
corresponding time.
For definiteness, we consider the scattering problem and assume that the
boundary points of the particle trajectories lie in the asymptotic
regions of free motion.
The squares of the velocity vectors and their scalar
product are implied in the Minkowski space with the signature $\left(
+.-,-,-\right) $. The same applies to the squared distance between particles
\begin{equation}
s_{12}^{2}=\left( x_{1}\left( s_{1}\right) -x_{2}\left( s_{2}\right) \right)
^{2}  \label{2}
\end{equation}%
The Fokker action Eq.(\ref{1}) contains two independent parameters of time,
which excludes the usual Hamiltonian form of the theory and the use of the
usual rules of canonical quantization. This was partly an incentive for the
development of a new formulation of quantum mechanics in terms of the
functional integral \cite{4}. This formulation turned out to be useful, in
particular, in the case of systems with a nonlocal in time interaction
potential \cite{5}. However, for the Fokker action Eq.(\ref{1}), quantum
theory in terms of the functional integral was not formulated. There are two
reasons for this. The first is the uncertainty of the integration measure in
the space of world particle lines. This issue is considered in this paper.
In the framework of the generalized canonical formulation of the Fokker
theory, a representation of the Feynman propagator for a system of
interacting particles will be  obtained. The second reason is common to
relativistic quantum mechanics: the Feynman propagator has no dynamic
meaning, since the proper time of particles is not an observable parameter.
This issue requires a separate consideration. For ordinary systems with
a local interaction potential, there is a canonical representation, and it
is the initial one in the definition of the functional integral in the phase
space for the kernel of the evolution operator - the solution of the Schr%
\"{o}dinger equation \cite{6}. From it, after integration over the momenta,
we obtain the Feynman integral in the configuration space. The canonical
form of action Eq.(\ref{1}) is also desirable for determining the functional
integral. Despite the multi-temporal nature of dynamics in the Fokker
theory, it admits a generalized canonical formulation in which the world
lines of particles as a whole serve as canonical coordinates \cite{7,8}.

In
this paper, a variant of such a generalization is considered. Based on it,
the functional integral on the generalized phase space is determined, which,
after integration over the momenta, becomes the functional integral on the
configuration space of world particle lines with a certain measure. The
construction of the Fokker quantum theory is being completed by integrating
particles in their own time. In the case of a single particle, the integral
of the kernel of the evolution operator over its proper time is a
representation of the Green function of the Dirac and Klein-Gordon
operators, i.e. the Feynman propagator \cite{9,10}. The result of multiple
Fock-Schwinger integration (FS) in the Fokker theory will also be called the
Feynman propagator. However, the Feynman propagator itself does not describe
the evolution of the system, but is a skeletal element of the diagram
technique in perturbation theory \cite{11}. To introduce the physical
parameters of time and the dynamic interpretation of the Feynman propagator,
a modification of the Fokker theory at the classical level is required,
similar to that considered in the case of a single relativistic particle in
\cite{12}.

The next section proposes a generalized canonical formalism for
multitemporal dinamic theory and defines the Feynman propagator in the
Fokker theory. In the third section, a modification of the Feynman
propagator is proposed, allowing its dynamic interpretation.

\section{GENERALIZED CANONICAL REPRESENTATION AND QUANTIZATION OF FOKKER
THEORY}

In the absence of the Lagrange function, we consider action Eq.(\ref{1}%
) as the velocity $\overset{\cdot }{x}_{1\mu }\left( s_{1}\right) ,x_{2\nu
}^{\prime }\left( s_{2}\right) $ functional, which should now be considered
as an infinite set of quantities, also numbered by the particle's own time,
and we define canonical momenta as variational derivatives:
\begin{eqnarray}
p_{1\mu }\left( s_{1}\right) &=&\frac{\delta I_{F}}{\delta \overset{\cdot }{x%
}_{1}^{\mu }\left( s_{1}\right) }=\overset{\cdot }{x}_{1\mu }\left(
s_{1}\right)  \notag \\
&&+e_{1}e_{2}\int_{0}^{S_{2}}ds_{2}\delta \left( s_{12}^{2}\right) x_{2\mu
}^{\prime }\left( s_{2}\right) ,  \label{3}
\end{eqnarray}
\begin{eqnarray}
p_{2\nu }\left( s_{2}\right) &=&\frac{\delta I_{F}}{\delta x_{2}^{\prime \nu
}\left( s_{2}\right) }=x_{2\nu }^{\prime }\left( s_{2}\right)  \notag \\
&&+e_{1}e_{2}\int_{0}^{S_{1}}ds_{1}\delta \left( s_{12}^{2}\right) \overset{%
\cdot }{x}_{1\nu }\left( s_{1}\right) .  \label{4}
\end{eqnarray}
Next, we act in a standard way. We introduce the generalized Hamiltonian
functional $H_{F}\left[ p_{1},x_{1},p_{2},x_{2}\right] $ by means of the
generalized Legendre transform:
\begin{eqnarray}
H_{F}\left[ p_{1},x_{1},p_{2},x_{2}\right] &=&\int_{0}^{S_{1}}ds_{1}p_{1\mu
}\left( s_{1}\right) \overset{\cdot }{x}_{1}^{\mu }\left( s_{1}\right)
\notag \\
&&+\int_{0}^{S_{2}}ds_{2}p_{2\nu }\left( s_{2}\right) x_{2}^{\prime \nu
}\left( s_{2}\right)  \notag \\
&&-I_{F}\left[ \overset{\cdot }{x}_{1},x_{1},x_{2}^{\prime },x_{2}\right] ,
\label{5}
\end{eqnarray}%
where velocities should also be excluded by solving Eqs.(\ref{3}) and (\ref{4}).
The solution can be written in the form of series of
perturbation theory:
\begin{equation}
\overset{\cdot }{x}_{1\mu }\left( s_{1}\right) =p_{1\mu }\left( s_{1}\right)
-e_{1}e_{2}\int_{0}^{S_{2}}ds_{2}\delta \left( s_{12}^{2}\right) p_{2\mu
}\left( s_{2}\right) +...,  \label{6}
\end{equation}
\begin{equation}
x_{2\nu }^{\prime }\left( s_{2}\right) =p_{2\nu }\left( s_{2}\right)
-e_{1}e_{2}\int_{0}^{S_{1}}ds_{1}\delta \left( s_{12}^{2}\right) p_{1\nu
}\left( s_{1}\right) +....  \label{7}
\end{equation}%
Substituting  Eqs. (\ref{6}) and (\ref{7}) into Eq.(\ref{5}), we obtain the
generalized Hamilton functional in the form of a quadratic form of
generalized momenta:
\begin{equation*}
H_{F}\left[ p_{1},x_{1},p_{2},x_{2}\right]
\end{equation*}
\begin{eqnarray}
&=&\sum\limits_{\alpha ,\beta =1,2}\int_{0}^{S_{\alpha }}ds_{\alpha
}\int_{0}^{S_{\beta }}ds_{\beta }M_{\alpha \mu \beta \nu }\left( s_{\alpha
},s_{\beta }\right)  \notag \\
&&\times p_{\alpha \mu }\left( s_{\alpha }\right) p_{\beta \nu }\left(
s_{\beta }\right) -m_{1}^{2}S_{1}-m_{2}^{2}S_{2},  \label{8}
\end{eqnarray}%
where $M_{\alpha \mu \beta \nu }\left( s_{\alpha },s_{\beta }\right) $ are
the matrix entries that are series composed of multiple integrals of the
products $\delta \left( s_{\alpha \beta }^{2}\right) $. They are explicitly
contained in the equations for the generalized canonical momenta Eqs. (\ref{6}%
) and (\ref{7}) that follow from the generalized canonical form of the
Fokker action,
\begin{eqnarray}
\widetilde{I}_{F}\left[ p_{1},x_{1},p_{2},x_{2}\right] &=&%
\int_{0}^{S_{1}}ds_{1}p_{1\mu }\left( s_{1}\right) \overset{\cdot }{x}%
_{1}^{\mu }\left( s_{1}\right)  \notag \\
&&+\int_{0}^{S_{2}}ds_{2}p_{2\nu }\left( s_{2}\right) x_{2}^{\prime \nu
}\left( s_{2}\right)  \notag \\
&&-H_{F}\left[ p_{1},x_{1},p_{2},x_{2}\right] .  \label{9}
\end{eqnarray}%
It is in this form the Fokker action is taken as the basis for
constructing the functional integral on the generalized phase space, which
determines the evolution of the system in terms of the particle's proper
time:
\begin{eqnarray}
&&K\left( S_{1},S_{2}\right)  \notag \\
&=&\int \frac{dp_{1out}}{2\pi \hbar }\prod\limits_{s_{1},\mu }\frac{dp_{1\mu
}dx_{1\mu }}{2\pi \hbar }\frac{dp_{2out}}{2\pi \hbar }\prod\limits_{s_{2},%
\nu }\frac{dp_{2\nu }dx_{2\nu }}{2\pi \hbar }  \notag \\
&&\times \exp \left\{ \frac{i}{\hbar }\widetilde{I}_{F}\left[
p_{1},x_{1},p_{2},x_{2}\right] \right\} .  \label{10}
\end{eqnarray}%
Gaussian momentum integrals are easily calculated, and we arrive at a
functional integral on the configuration space with the initial Fokker
action:
\begin{eqnarray}
K\left( S_{1},S_{2}\right) &=&\int \prod\limits_{s_{1},\mu }dx_{1\mu
}\prod\limits_{s_{2},\nu }dx_{2\nu }\frac{1}{\sqrt{\det M}}  \notag \\
&&\times \exp \left\{ \frac{i}{\hbar }I_{F}\left[ \overset{\cdot }{x}%
_{1},x_{1},x_{2}^{\prime },x_{2}\right] \right\} .  \label{11}
\end{eqnarray}%
Thus, a measure of functional integration in the configuration space of
particle trajectories is defined. This integral can be made convenient for
calculations in the framework of perturbation theory by introducing the
determinant of the inverse matrix $A=M^{-1}$,
\begin{equation}
\sqrt{\det A}=\frac{1}{\sqrt{\det M}},  \label{12}
\end{equation}%
and take into account that it is determined by the coefficients of
Eqs.(\ref{3}) and (\ref{4}). Then, for the
determinant, we obtain the integral
representation using Grassmann variables \cite{6}:
\begin{eqnarray}
\det A &=&\int \prod\limits_{s_{1},\mu }\frac{d\chi _{1\mu }^{\ast }d\chi
_{1\mu }}{2\pi }\prod\limits_{s_{2},\nu }\frac{d\chi _{2\nu }^{\ast }d\chi
_{2\nu }}{2\pi }  \notag \\
&&\times \exp \left[ \int_{0}^{S_{1}}ds_{1}\chi _{1}^{\ast }\chi
_{1}+\int_{0}^{S_{2}}ds_{2}\chi _{2}^{\ast }\chi _{2}\right.  \notag \\
&&\left. +e_{1}e_{2}\int_{0}^{S_{1}}ds_{1}\int_{0}^{S_{2}}ds_{2}\delta
\left( s_{12}^{2}\right) \right.  \notag \\
&&\left. \times \left( \chi _{1}^{\ast }\chi _{2}+\chi _{2}^{\ast }\chi
_{1}\right) \right] .  \label{13}
\end{eqnarray}

The functional integral Eq. (\ref{11}) is a function of the boundary points of
the world lines of particles at extreme times:
$K\left( S_{1},x_{1out},0,x_{1in}S_{2},x_{2out},0,x_{2in}\right) $.
Now it is necessary to make the next step that will complicate the dynamic
interpretation of the Fokker quantum theory. Since the internal time of
particles in quantum theory is considered unobservable, we obtain the
covariant Feynman propagator by integrating the FS along these parameters:
\begin{equation}
\mathit{K}\left( x_{1out},x_{1in};x_{2out},x_{2in}\right) =\int_{0}^{\infty
}dS_{1}\int_{0}^{\infty }dS_{2}K\left( S_{1},S_{2}\right) .  \label{14}
\end{equation}%
In the absence of interaction, propagator Eq.(\ref{14}) is reduced to the
product of ordinary Feynman propagators of scalar particles.

\section{MODIFIED FEYNMAN PROPAGATOR OF FOKKER THEORY}

Having thus determined the covariant Feynman propagator in the Fokker
quantum theory, one should think about its dynamic interpretation. This will
require its modification. According to \cite{12}, the modification of the
action is carried out at the classical level: its variation, generated by
the infinitesimal shift of the proper time in the coordinates $x_{10}\left(
s_{1}\right) ,x_{20}\left( s_{2}\right) $ of Minkowski space,
\begin{equation}
\delta x_{10}=\overset{\cdot }{x}_{10}\varepsilon _{1},\delta
x_{20}=x_{20}^{\prime }\varepsilon _{2}.  \label{15}
\end{equation}
is added to it. One can show that, by analogy with \cite{12}, the
modified Feynman propagator for interacting charges in the Fokker theory has
the form:
\begin{eqnarray}
&&\widetilde{\mathit{K}}\left(
P_{1out},x_{1out},P_{1in},x_{1in};P_{2out},x_{2out},P_{2in},x_{2in}\right)
\notag \\
&=&\int_{0}^{\infty }dS_{1}\int_{0}^{\infty }dS_{2}\int
\prod\limits_{s_{1},k}dx_{1k}\prod\limits_{s_{2},l}dx_{2l}\prod%
\limits_{s_{1}}dP_{1}\prod\limits_{s_{2}}dP_{2}  \notag \\
&&\times \sqrt{\det \widetilde{A}}\prod\limits_{s_{1}}\delta \left( \overset{%
\cdot }{x}_{10}-\sqrt{2P_{1}}\right) \prod\limits_{s_{2}}\delta \left(
x_{20}^{\prime }-\sqrt{2P_{2}}\right)  \notag \\
&&\times \prod\limits_{s_{1}}\delta \left( \overset{\cdot }{P}_{1}+\sqrt{%
2P_{1}}\overset{\cdot }{x}_{1k}F_{1k}\right)  \notag \\
&&\times \prod\limits_{s_{2}}\delta \left( P_{2}^{\prime }-\sqrt{2P_{2}}%
x_{2l}^{\prime }F_{2l}\right)  \notag \\
&&\times \exp \left\{ \frac{i}{\hbar }\widetilde{I}_{F}\left[ \overset{\cdot
}{x}_{1},x_{1},P_{1},x_{2}^{\prime },x_{2},P_{2}\right] \right\} .
\label{16}
\end{eqnarray}%
Here $P_{1},P_{2}$ are the self-energies of the particles, and
\begin{eqnarray}
F_{1k} &=&2e_{1}e_{2}\int_{0}^{\infty }dS_{2}\frac{d}{ds_{12}^{2}}\delta
\left( s_{12}^{2}\right)  \notag \\
&&\times \left[ \left( x_{10}-x_{20}\right) x_{2k}^{\prime }-\left(
x_{1k}-x_{2k}\right) x_{20}^{\prime }\right] ,  \label{17}
\end{eqnarray}
\begin{eqnarray}
F_{2l} &=&2e_{1}e_{2}\int_{0}^{\infty }dS_{1}\frac{d}{ds_{12}^{2}}\delta
\left( s_{12}^{2}\right)  \notag \\
&&\times \left[ \left( x_{10}-x_{20}\right) \overset{\cdot }{x}_{1l}-\left(
x_{1k}-x_{2k}\right) \overset{\cdot }{x}_{10}\right]  \label{18}
\end{eqnarray}%
are forces acting on each particle,
\begin{eqnarray}
&&\widetilde{I}_{F}\left[ \overset{\cdot }{x}_{1},x_{1},P_{1},x_{2}^{\prime
},x_{2},P_{2}\right]  \notag \\
&=&\frac{1}{2}\int_{0}^{S_{1}}ds_{1}\left( 2P_{1}-\overset{\cdot }{x}%
_{1k}^{2}+m_{1}^{2}\right)  \notag \\
&&+\frac{1}{2}\int_{0}^{S_{2}}ds_{2}\left( 2P_{2}-\overset{\cdot }{x}%
_{2l}^{2}+m_{2}^{2}\right)  \notag \\
&&+e_{1}e_{2}\int_{0}^{S_{1}}ds_{1}\int_{0}^{S_{2}}ds_{2}\delta \left(
s_{12}^{2}\right)  \notag \\
&&\times \left[ \sqrt{2P_{1}2P_{2}}-\overset{\cdot }{x}_{1k}x_{2k}^{\prime }%
\right]  \label{19}
\end{eqnarray}%
is a modified Fokker action, and
\begin{eqnarray}
\det \widetilde{A} &=&\int \prod\limits_{s_{1},k}\frac{d\chi _{1k}^{\ast
}d\chi _{1k}}{2\pi }\prod\limits_{s_{2},l}\frac{d\chi _{2l}^{\ast }d\chi
_{2l}}{2\pi }  \notag \\
&&\times \exp \left[ \int_{0}^{S_{1}}ds_{1}\chi _{1k}^{\ast }\chi
_{1k}+\int_{0}^{S_{2}}ds_{2}\chi _{2l}^{\ast }\chi _{2l}\right.  \notag \\
&&\left. +e_{1}e_{2}\int_{0}^{S_{1}}ds_{1}\int_{0}^{S_{2}}ds_{2}\delta
\left( s_{12}^{2}\right) \right.  \notag \\
&&\left. \times \left( \chi _{1k}^{\ast }\chi _{2k}+\chi _{2k}^{\ast }\chi
_{1k}\right) \right]  \label{20}
\end{eqnarray}%
is a measure of integration in the space of particle trajectories. Here we
turned to three-dimensional notation.

The presence of $\delta $-functions in the functional integral Eq. (\ref{16})
makes it possible to remove the FS integration over the particle proper time
after solving the classical equations of motion for $\left(
x_{10},P_{1}\right) $ and $\left( x_{20},P_{2}\right) $. In this case, the
role of time for each particle, as one would expect, is assumed by their $%
x_{10},x_{20}$ - coordinates in Minkowski space, and the self-energies of
particles $P_{1},P_{2}$ are additional observables.

\section{CONCLUSIONS}

Relativistic quantum mechanics of a long-range interaction particle system
is formulated in terms of a functional integral on the generalized phase
space of world lines parameterized by the particle's proper time. The
covariance of quantum theory is ensured by integration over the parameters
of proper time in the Feynman propagator. A dynamic interpretation of the
covariant quantum theory can be achieved by a modification of Fokker theory
at the classical level. This modification does not change the dynamic
content of the classical theory, but allows to remove the integration
over proper time of particles in the Feynman propagator. As a result, the
role of independent time parameters is taken by the $x_{10},x_{20}$
coordinates of the particles in Minkowski space.

\section{ACKNOWLEDGEMENTS}

We thank V.A. Franke for useful discussions.




\end{document}